\pdfoutput=1 
\documentclass[lettersize,journal]{IEEEtran}
\usepackage{amsmath,amsfonts}
\usepackage{algorithmic}
\usepackage{algorithm}
\usepackage{array}
\usepackage{textcomp}
\usepackage{stfloats}
\usepackage{url}
\usepackage{verbatim}
\usepackage{graphicx}
\usepackage{cite}
\usepackage{multirow}
\usepackage{makecell}
\usepackage{pifont}
\usepackage{caption}
\usepackage{subcaption}
\usepackage[most]{tcolorbox}
\usepackage{colortbl,hhline}
\usepackage[normalem]{ulem}
\usepackage{float}
\usepackage{fontawesome}
\usepackage{subfiles}
\usepackage{booktabs}
\usepackage{bm}
\usepackage{fontawesome}
\usepackage{threeparttable}
\usepackage{tabularx}

\newcolumntype{P}[1]{>{\centering\arraybackslash}p{#1}}

\usepackage[colorlinks,linkcolor=blue,anchorcolor=blue,citecolor=blue]{hyperref}

\begin{document}

\title{When Models Meet Users: An Empirical Study of Perceptions of General LLMs and Multimodal LLMs on Hugging Face}

\author{Yujian Liu, Xiao Yu, Jacky Wai Keung, Xing Hu, Xin Xia, Xiaoxue Ma
\thanks{Yujian Liu, Xiao Yu, Xing Hu, and Xin Xia are with the State Key Laboratory of Blockchain and Data Security, Zhejiang University, Hangzhou, China.  E-mail: liuyujian@zju.edu.cn, xiao.yu@zju.edu.cn, xinghu@zju.edu.cn, and xin.xia@acm.org. Jacky Wai Keung is with Department of Computer Science, City University of Hong Kong, Hong Kong, China. E-mail: jacky.keung@cityu.edu.hk. Xiaoxue Ma is with Department of Electronic Engineering and Computer Science, Hong Kong Metropolitan University, Hong Kong, China. Email: kxma@hkmu.edu.hk.}
}

\maketitle

\begin{abstract}

Large language models (LLMs) have rapidly evolved from general-purpose systems to multimodal models capable of processing text, images, and audio. As both general-purpose LLMs (GLLMs) and multimodal LLMs (MLLMs) gain widespread adoption, understanding user perceptions in real-world settings becomes increasingly important. However, existing studies often rely on surveys or platform-specific data (e.g., Reddit or GitHub issues), which either constrain user feedback through predefined questions or overemphasize failure-driven, debugging-oriented discussions, thus failing to capture diverse, experience-driven, and cross-model user perspectives in practice. 
To address this issue, we conduct an empirical study of user discussions on Hugging Face, a major model hub with diverse models and active communities. We collect and manually annotate 662 discussion threads from 38 representative models (21 GLLMs and 17 MLLMs), and develop a three-level taxonomy to systematically characterize user concerns. 
Our analysis reveals that LLM access barriers, generation quality, and deployment and invocation complexity are the most prominent concerns, alongside issues such as documentation limitations and resource constraints. Based on these findings, we derive actionable implications for improving LLM ecosystem. 

\end{abstract}

\begin{IEEEkeywords}
Large Language Models, Multimodal Large Language Models, User Perception, Empirical Study,  Hugging Face
\end{IEEEkeywords}

\section{Introduction}

In recent years, general-purpose large language models (GLLMs) have been widely adopted across a broad range of applications, including natural language understanding~\cite{bubeck2023paper,chowdhery2023palm}, content generation~\cite{geng2024large, xue2024automated}, and software development assistance~\cite{ma2025specgen, huang2025comprehensive}. Meanwhile, multimodal large language models (MLLMs), which can process and integrate multiple modalities such as text, images, and audio, have rapidly emerged and attracted increasing attention ~\cite{yin2024survey, caffagni2024revolution}.
As both GLLMs and MLLMs gain widespread adoption, understanding how users perceive these models in practice becomes increasingly important. User perceptions reflect real-world experiences regarding model capabilities, usability, limitations, and expectations, and can provide valuable insights to guide the design and development of future LLM systems. 

Despite the growing body of research on user perceptions and experiences with LLMs, existing studies present several limitations. Prior work has explored user perceptions through online Reddit discussions~\cite{katta2025analyzing} and surveys~\cite{wang2024understanding}, as well as domain-specific contexts such as education~\cite{bernabei2023students}, healthcare~\cite{choudhury2025user}, and LLM-based coding assistants~\cite{github2024survey,liang2024large,sergeyuk2025using,wang2023practitioners,mcnutt2023design,ziegler2022productivity,lyu2025my}. However, these studies primarily focus on end users and some rely on predefined survey instruments, which may limit the diversity and authenticity of real-world feedback. 
Complementary research has investigated challenges in LLM development and deployment. For instance, Yu et al.~\cite{yu2026does} analyzed GitHub issues of DeepSeek, LLaMA, and Qwen to identify failure symptoms and root causes in fine-tuning and deployment, while Chen et al.~\cite{chen2025empirical} examined challenges faced by developers building applications with GPT models based on OpenAI developer forum data.  However, these studies typically focus on a limited set of models and predominantly capture failure-driven, problem-centric discussions, mainly reflecting debugging and troubleshooting processes. Consequently, there remains a limited understanding of how practitioners perceive and experience modern LLMs in real-world usage across diverse models, tasks, and deployment scenarios.

To address these gaps, this paper presents an in-depth empirical study of user discussions on Hugging Face, covering both GLLMs and MLLMs. Unlike traditional code-hosting platforms or general forums, Hugging Face serves as the central hub for the modern AI ecosystem, bridging the gap between model developers and downstream practitioners. It hosts a wide variety of models alongside active community discussions where users report not only technical bugs but also share deployment configurations, prompt engineering experiences, and practical adoption barriers. 
By systematically collecting and manually analyzing representative discussion threads across both GLLMs and MLLMs, we develop a three-level taxonomy to systematically categorize user concerns into seven dimensions: \textit{functionality}, \textit{usability}, \textit{reliability}, \textit{supportability}, \textit{performance}, \textit{comparison}, and \textit{general experience}.

Overall, our analysis reveals several key insights into user perceptions of GLLMs and MLLMs, highlighting the current maturity, bottlenecks, and real-world dynamics of the LLM ecosystem. Access barriers, particularly in the Llama model family, remain a major obstacle for practical adoption, showing that model openness does not necessarily translate into accessibility. Model invocation and deployment represent the most common engineering challenges, especially for MLLMs with more heterogeneous architectures, and are further exacerbated by insufficient documentation and developer support. Generation quality remains a central factor shaping user perceptions, with MLLM users being particularly sensitive to output quality and prompt design. Although benchmark evaluations remain important for model comparison, difficulties in reproducing reported results raise concerns about the reliability and interpretability of benchmark-based evaluations. Fine-tuning is widely adopted for customization, particularly for smaller models, yet remains technically challenging even with parameter-efficient methods such as LoRA. Meanwhile, GPU and memory constraints continue to limit local deployment, motivating the widespread adoption of quantization techniques. These findings suggest that future LLM development should move beyond capability optimization toward usability-oriented engineering, including transparent access mechanisms, standardized deployment practices, task-oriented documentation, efficient adaptation support, and reliable evaluation frameworks. 

This work makes the following key contributions:

(1) We conduct a large-scale empirical study of 662 manually annotated discussion threads from Hugging Face to characterize the open-source LLM ecosystem, covering both GLLMs and MLLMs, and develop a three-level taxonomy consisting of 7 categories, 17 subcategories, and 58 leaf nodes.

 (2) We identify several previously underexplored aspects of real-world LLM usage, including access and licensing barriers, the role of prompt design in shaping perceived generation quality, invocation and deployment challenges, the concentration of fine-tuning on smaller models, benchmark reproducibility difficulties, documentation-related learning costs, and multilingual support needs. Based on these findings, we further derive actionable implications for improving the LLM ecosystem.

\section{Methodology}
\label{sec:methodology}

\begin{figure}[!t]
    \centering
    \includegraphics[width=\linewidth]{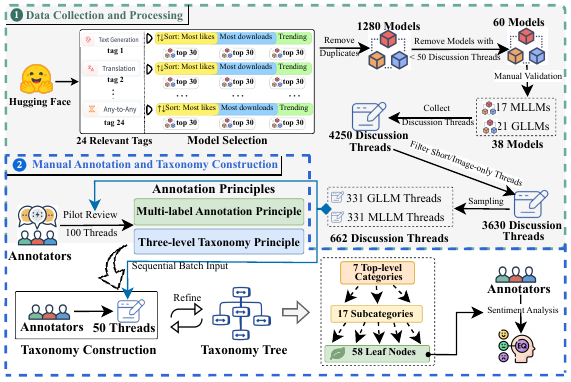}
    \captionsetup{skip=3pt} 
    \caption{The overview of the research methodology.}
    \label{fig:overview}
\end{figure}

Figure~\ref{fig:overview} shows an overview of our research methodology.

\subsection{Data Collection and Processing} 
\noindent \textbf{Model Selection.}
Hugging Face has become the largest open-source platform for hosting deep learning models, pretrained models, and LLMs, providing millions of models across diverse tasks. 
To identify candidate LLMs for this study, we first leverage the task tags provided by the platform. Based on the technical characteristics of GLLMs and MLLMs, we select 24 relevant tags (shown in Table~\ref{tab:tags}) covering both textual tasks (e.g., \textit{text-generation}, \textit{translation}, and \textit{question-answering}) and multimodal tasks (e.g., \textit{text-to-image} and \textit{text-to-video}). This strategy ensures that the candidate pool includes both GLLMs and emerging MLLMs.  
For each selected tag, models are ranked according to three indicators available on Hugging Face: \textit{likes}, \textit{downloads}, and \textit{trending}. We collect the top 30 models for each metric to capture both long-term popularity (\textit{likes} and \textit{downloads}) and short-term community attention (\textit{trending}). After removing duplicates across tags and metrics, we obtain an initial candidate set of 1,280 unique models.

\noindent \textbf{Activity Filtering and Manual Validation.}
Hugging Face provides discussion forums where users report issues, share experiences, and request features, offering rich insights into real-world usage. After filtering out pull requests and system messages to retain only user-generated threads, we apply an activity threshold of 50 threads. This threshold strikes a necessary balance: a higher limit excessively narrows the scope of models, while a lower one lacks the active engagement required to yield valuable insights into user perceptions. Ultimately, this reduces our candidate pool from 1,280 to 60 active models.
We then manually verify whether the remaining models correspond to GLLMs or MLLMs. Models outside the scope of our study, such as model plugins (e.g., \textit{h94/IP-Adapter-FaceID}~\cite{ip_adapter_faceid}), as well as models whose discussions are dominated by non-technical or politically oriented debates (e.g., \textit{perplexity-ai/r1-1776}), are excluded. After this validation process, we obtain a final set of 38 models, including 21 GLLMs from 9 model families and 17 MLLMs from 13 model families (shown in Table~\ref{tab:models}).

\noindent \textbf{Sampling.}
Using the Hugging Face API, we collect all discussion threads associated with these models, treating each thread as one record. Each discussion thread consists of an initial post and may include multiple user replies; we treat the entire thread as a single analysis unit. 
In total, we obtain 4,250 records (2,808 for GLLMs and 1,442 for MLLMs). Non-English content is first translated into English using Google Translate, and extremely short posts (containing fewer than 10 words) as well as image-only threads are then removed. After cleaning, 3,630 valid discussion threads remain, including 2,383 for GLLMs and 1,247 for MLLMs. Given the relatively large dataset, we apply statistical sampling to facilitate qualitative analysis, ensuring that the sampled threads are representative, manageable, and suitable for drawing reliable insights. Following prior empirical studies~\cite{hata20199,lyu2024evaluating,lyu2025my}, we adopt a sampling strategy with a 95\% confidence level and a 5\% margin of error. From the GLLM discussion pool, we randomly sample 331 threads. To enable a balanced comparison between model types, we also randomly sample 331 threads from the MLLM discussions. 

\begin{table}[!t]
    \centering
    \fontsize{8pt}{8pt}\selectfont
    \caption{The set of task tags used for model retrieval.}
    \label{tab:tags}
    \begin{tabular}{@{}p{0.7cm}<{\centering}|p{7.4cm}@{}}
    \toprule
        \textbf{Type} & \textbf{Tags}  \\ \midrule

         \textbf{GLLM} & text-generation, translation, table-question-answering \newline
         question-answering, summarization \\ \midrule
         \textbf{MLLM}& text-to-image, text-to-video,image-to-text \newline
         image-to-image,image-to-video, visual-question-answering \newline
         unconditional-image-generation, mask-generation \newline
         text-to-3d, image-to-3d, video-to-video \newline
         document-question-answering, visual-document-retrieval \newline
         audio-text-to-text, image-text-to-text, image-text-to-image \newline
         image-text-to-video, video-text-to-text, any-to-any \\
        \bottomrule
    \end{tabular}
\end{table}

\begin{table}[!t]
    \centering
    \fontsize{8pt}{8pt}\selectfont
    \caption{The list of selected general and multimodal LLMs.}
    \label{tab:models}
    \begin{tabular}{@{}p{0.7cm}<{\centering}|p{7.4cm}@{}}
    \toprule
        \textbf{Type} & \textbf{Model Name}  \\ \midrule

         \textbf{GLLM} & deepseek-ai/DeepSeek-R1, meta-llama/Meta-Llama-3-8B \newline
         meta-llama/Llama-3.1-8B-Instruct, openai/gpt-oss-120b \newline
         meta-llama/Meta-Llama-3-8B-Instruct, openai/gpt-oss-20b \newline
         mistralai/Mistral-7B-v0.1, deepseek-ai/DeepSeek-V3 \newline
         microsoft/phi-2, google/gemma-7b, bigcode/starcoder\newline
         deepseek-ai/DeepSeek-V3-0324, tiiuae/falcon-40b \newline
         mistralai/Mistral-7B-Instruct-v0.2, Qwen/QwQ-32B \newline
         meta-llama/Llama-3.3-70B-Instruct, meta-llama/Llama-3.2-1B \newline
         deepseek-ai/DeepSeek-R1-0528, meta-llama/Llama-3.1-8B \newline
         nvidia/Llama-3.1-Nemotron-70B-Instruct-HF \newline
         meta-llama/Llama-3.2-1B-Instruct
         \\ \midrule
         \textbf{MLLM}& Phr00t/Qwen-Image-Edit-Rapid-AIO, deepseek-ai/Janus-Pro-7B \newline
         meta-llama/Llama-4-Scout-17B-16E-Instruct, adept/fuyu-8b \newline
         microsoft/Phi-4-multimodal-instruct, google/gemma-3-27b-it \newline
         microsoft/Florence-2-large, PaddlePaddle/PaddleOCR-VL \newline
         meta-llama/Llama-3.2-11B-Vision-Instruct \newline
         Qwen/Qwen2.5-VL-7B-Instruct, Qwen/Qwen2-VL-7B-Instruct \newline
         google/gemma-3-4b-it, HuggingFaceM4/idefics2-8b \newline
         llava-hf/llava-1.5-7b-hf, deepseek-ai/DeepSeek-OCR \newline
         openbmb/MiniCPM-Llama3-V-2\_5, vikhyatk/moondream2 \\
        \bottomrule
    \end{tabular}
\end{table}

\subsection{Manual Annotation and Taxonomy Construction}

\noindent \textbf{Initial Discussion and Coding Principles.}
To establish annotation principles, two annotators with over three years of experience working on LLMs review a random 100-record pilot sample. We analyze both text and images to identify recurring themes, ensuring a comprehensive understanding of the dataset's characteristics. This process leads to two key principles:
(1) \textit{Multi-label annotation principle}. A single discussion thread may contain multiple independent topics (e.g., hardware requirements followed by comments on generation quality). Therefore, when multiple distinct themes appear, multiple labels can be assigned to ensure complete topic coverage. 
(2) \textit{Three-level taxonomy principle}. During the pilot review, we observe that a simple two-level structure (category–leaf node) is too coarse to capture the diversity of user concerns. Following prior work~\cite{humbatova2020taxonomy,lyu2025my}, we organize the taxonomy into three levels—\textit{category}, \textit{subcategory}, and \textit{leaf node}—to represent discussion topics with finer granularity. 
Based on insights from the pilot review and existing taxonomies in app review mining~\cite{kurtanovic2017mining} and recent studies on LLM-based coding assistant reviews~\cite{lyu2025my}, we also establish an initial set of top-level categories, including \textit{Functionality}, \textit{Usability}, \textit{Supportability}, \textit{Performance}, and \textit{General Experience}.

\noindent \textbf{Bottom-Up Taxonomy Construction.}
Building upon the predefined top-level categories and annotation principles, we construct the taxonomy through a bottom-up process. The coding is conducted in sequential batches of 50 discussion threads.
In each batch, the two annotators independently assign one or more labels to each thread based on its discussed topics. They then compare their labeling results and resolve any disagreements through discussion with an additional researcher. For threads that cannot be assigned to any existing leaf node at the time of annotation, the annotators temporarily label them as ``unknown.'' These cases are later jointly reviewed by the annotation team to determine whether they should be merged into existing categories or used to create new leaf nodes, enabling incremental expansion of the taxonomy. Semantically similar labels or different expressions referring to the same concept are merged and normalized into unified leaf nodes through consensus within the annotation team.
Initially, these leaf nodes are directly grouped under the predefined top-level categories, forming a two-level structure. When a category accumulates more than four leaf nodes with clear thematic differences, an intermediate \textit{subcategory} layer is inserted between the category and leaf nodes, resulting in a three-level taxonomy (category–subcategory–leaf node). The taxonomy is gradually refined as we proceed through the labeling batches.

\noindent \textbf{Taxonomy Refinement and Final Annotation.} 
After completing the first four batches (200 threads in total), we identify gaps in the taxonomy: several threads cannot be clearly categorized under the existing top-level categories. To address this issue, we introduce two additional top-level categories—\textit{Reliability} and \textit{Comparison}—and revisit the previously annotated threads to reassign them to the appropriate categories.
Following this refinement, the remaining 462 threads ($=662-50\times4$) are independently labeled by the two annotators. Any remaining disagreements are resolved through discussion with the additional researcher. In the final three annotation rounds, no discussion threads are labeled as ``unknown,'' leading us to conclude that the taxonomy had reached saturation. The inter-rater agreement for this final independently labeled phase (before discussion with the additional researcher) is Cohen's $\kappa=0.87$, indicating a relatively high level of consistency between the two annotators.
In total, we manually annotate 662 discussion threads, resulting in 58 unique leaf nodes organized into 17 subcategories and 7 top-level categories.

\noindent \textbf{Sentiment Annotation.}
A single discussion thread may express multiple sentiments across different discussion topics. Therefore, instead of assigning one sentiment to each discussion thread, we conduct sentiment analysis at the leaf-node level. Each labeled leaf node is marked as \textit{like}, \textit{dislike}, or \textit{neutral} (including feature requests or emotionally neutral mentions).
We define \textit{neutral} as the default label when no explicit positive or negative sentiment is expressed; \textit{like} and \textit{dislike} are assigned only when sentiment is explicitly stated, in order to reduce subjectivity in interpretation.
The two annotators first independently annotate 20\% of the data to establish a common understanding of the sentiment categories. Disagreements are resolved through discussion with the additional researcher to reach a consensus. The remaining 80\% of the data is then annotated independently by the two annotators, with remaining disagreements again resolved through discussion with the additional researcher. The inter-rater agreement for the final independently labeled phase (before any discussion with the additional researcher) reaches Cohen's $\kappa=0.83$, indicating high consistency between the two annotators.

\section{Results}
\label{sec:results}

\begin{table*}[t]
    \caption{The categorization of discussion topics for general and multimodal LLMs. }
    \label{tab:category}
    \centering
    \resizebox{\textwidth}{!}{%
    \begin{tabular}{@{}l l | l | l | l | l | l | l@{}}
    \toprule
    \textbf{ID} & \textbf{Category-Subcategory }& \textbf{Description} & \textbf{GN} & \textbf{GR}  & \textbf{MN} & \textbf{MR} & \textbf{Total} \\
    \midrule
    
    \multirow{5}{*}{1} & \textbf{Functionality} & Model capabilities and task execution.  & 51 & 12.8\% & 84 & 22.0\% & 135 \\
    ~ & \quad $\bullet$ Core Capabilities  & Performance and fundamental capabilities on core tasks. & 20 & 5.0\% & 52 & 13.6\% & 72 \\
    ~ & \quad $\bullet$ Other Capabilities   & Requirements and discussions on specific functions.  & 8 & 2.0\% & 15 & 3.9\% & 23 \\
    ~ & \quad $\bullet$ Instruction Following   & Prompt usage, understanding, and output controllability. & 16 & 4.0\% & 14 & 3.7\% & 30 \\
    ~ & \quad $\bullet$ Extension   & External tool usage and module integration. & 7 & 1.8\% & 3 & 0.8\% & 10 \\
    \midrule
    
    \multirow{5}{*}{2} & \textbf{Usability} & Model usage, deployment, and development.  & 219 & 54.9\% & 178 & 46.6\% & 397 \\
    ~ & \quad $\bullet$ Documentation \& Tutorials   & Quality of documentation and example code. & 29 & 7.3\% & 43 & 11.3\% & 72 \\
    ~ & \quad $\bullet$ Invocation \& Deployment   & Model invocation and deployment experience.  & 46 & 11.5\% & 75 & 19.6\% & 121 \\
    ~ & \quad $\bullet$  Training \& Fine-tuning  & Task-specific adaptation and optimization. & 31 & 7.8\% & 36 & 9.4\% & 67 \\
    ~ & \quad $\bullet$ Access \& Licensing   & Model access control, licensing, and citation. & 113 & 28.3\% & 24 & 6.3\% & 137 \\
    \midrule
    
    \multirow{3}{*}{3} & \textbf{Reliability} & Model robustness and reliability during usage.  & 24 & 6.0\% & 16 & 4.2\% & 40 \\
    ~ & \quad $\bullet$ Bias \& Moderation  & Model bias and content moderation mechanisms. & 3 & 0.8\% & 1 & 0.3\% & 4 \\
    ~ & \quad $\bullet$ Bugs \& Errors  & System defects and abnormal model behaviors.  & 21 & 5.3\% & 15 & 3.9\% & 36 \\
    \midrule
    
    \multirow{3}{*}{4} & \textbf{Supportability} & External platform integration and dependency compatibility.  & 20 & 5.0\% & 28 & 7.3\% & 48 \\
    ~ & \quad $\bullet$ Third-party Integration  & Integration with external platforms and frameworks. & 7 & 1.8\% & 11 & 2.9\% & 18 \\
    ~ & \quad $\bullet$ Library Dependencies  & Compatibility with underlying libraries and dependencies.  & 13 & 3.3\% & 17 & 4.5\% & 30 \\
    \midrule
    
    \multirow{4}{*}{5} & \textbf{Performance} & Runtime performance, resource requirements, and efficiency.  & 42 & 10.5\% & 39 & 10.2\% & 81 \\
    ~ & \quad $\bullet$ Resource Consumption  & Consumption of computational resources and power. & 26 & 6.5\% & 21 & 5.5\% & 47 \\
    ~ & \quad $\bullet$ Inference Performance  & Generation speed and batch processing capabilities.  & 6 & 1.5\% & 4 & 1.0\% & 10 \\
    ~ & \quad $\bullet$ Model Efficiency & Quantization techniques and architectural lightweighting.  & 10 & 2.5\% & 14 & 3.7\% & 24 \\
    \midrule
    
    \multirow{3}{*}{6} & \textbf{Comparison} & Evaluations on benchmarks and cross-model comparisons.  & 13 & 3.3\% & 16 & 4.2\% & 29 \\
    ~ & \quad $\bullet$ Benchmarking & Quantitative performance on standardized benchmarks. & 8 & 2.0\% & 4 & 1.0\% & 12 \\
    ~ & \quad $\bullet$ Model Comparison & Comparison and evaluation of different models. & 5 & 1.3\% & 12 & 3.1\% & 17 \\
    \midrule
    
    \multirow{1}{*}{7} & \textbf{General Experience} & Overall subjective user feedback and sharing. & 30 & 7.5\% & 21 & 5.5\% & 51 \\ 
    \bottomrule
    \end{tabular}
    }
\end{table*}

\subsection{RQ1: What topics do users discuss about LLMs?}

Table~\ref{tab:category} presents the taxonomy derived from our analysis of Hugging Face discussion threads. The table headers are defined as follows: \textit{GN} and \textit{GR} denote the number and rate of threads associated with a given category or subcategory for GLLMs, respectively; \textit{MN} and \textit{MR} denote the number and rate for MLLMs; and \textit{Total} refers to the overall count across both types. 

\subsubsection{\textbf{Functionality}}

This category captures user discussions on model capabilities and task execution. It includes four subcategories: \textit{Core Capabilities}, \textit{Other Capabilities}, \textit{Instruction Following}, and \textit{Extension}, comprising 16 leaf nodes in total.

(1) \textit{\textbf{Core Capabilities}} focus on users’ feedback on performance and fundamental capabilities on core tasks. 
\textit{Generation quality} (53$\times$ \footnote{The multiplication symbol ($\times$) indicates the annotation frequency of each leaf node.}) is the most frequently discussed topic, referring to cases where outputs appear plausible or technically sound, and may even be correct, but still fail to meet user expectations, sometimes due to hallucinations. 
Discussions from MLLM users (46$\times$) substantially outnumber those from GLLM users (7$\times$). 
\textit{Input limitations} (10$\times$) concern constraints on model inputs, such as context window size. 
\textit{Reasoning ability} (7$\times$) reflects users’ evaluation of logical coherence during generation; all such discussions originate from GLLM users, as reasoning processes in MLLMs are less observable. 
Finally, \textit{Multi-turn dialogue} (2$\times$) addresses the model’s ability to maintain context and understand prior interactions.

(2) \textit{\textbf{Other Capabilities}} capture user requirements and discussions on specific functionalities. \textit{Language support} (7$\times$) is the most common request, indicating demand for specific languages (e.g., Japanese and French). 
Users also request features such as \textit{embedding output} (4$\times$), i.e., generating vector representations for downstream tasks, and \textit{batch processing} (4$\times$), such as processing multiple inputs (e.g., image or video sequences) in a single run. 
For MLLMs, users further expect multimodal capabilities, including uploading \textit{images} (1$\times$) and \textit{PDF files} (1$\times$), as well as functionalities such as \textit{image classification} (1$\times$), \textit{object detection} (4$\times$), and \textit{GUI support} (1$\times$). 

(3) \textit{\textbf{Instruction Following}} refers to discussions on prompt understanding and output controllability. \textit{Prompt and template usage} (23$\times$) is the dominant topic, covering how users design prompts or reusable templates to guide model behavior. 
\textit{Output control} (7$\times$) reflects users’ efforts to constrain model outputs to meet specific requirements, such as enforcing structured output formats (e.g., valid JSON).

(4) \textit{\textbf{Extension}} involves discussions on external tool usage and module integration. \textit{Tool usage} (7$\times$) is discussed exclusively by GLLM users and focuses on invoking external tool functions (e.g., calling a weather tool function) and related issues. 
In contrast, \textit{Module integration} (3$\times$) is only mentioned by MLLM users, reflecting the need to integrate models as modular components within larger systems (e.g., integrating a vision model as a CLIP \footnote{CLIP (Contrastive Language–Image Pretraining) is a model that learns joint representations of images and text, commonly used for cross-modal alignment.}-like module).

\subsubsection{\textbf{Usability}} 
This category captures user discussions on practical model usage, deployment, and development. It includes four subcategories: \textit{Documentation and Tutorials}, \textit{Invocation and Deployment}, \textit{Training and Fine-tuning}, and \textit{Access and Licensing}, comprising 12 leaf nodes.

(1) \textit{\textbf{Documentation \& Tutorials}} focuses on the quality and completeness of documentation and example code. 
\textit{Documentation} (43$\times$) includes both user requests for usage guidance and inquiries about basic model information. 
\textit{Example code} (29$\times$) covers shared code snippets as well as issues encountered when using them. 
In addition, users report errors and inconsistencies in both documentation and example code.

(2) \textit{\textbf{Invocation \& Deployment}} captures users’ practical experiences with invoking and deploying models. 
\textit{Model invocation} (61$\times$) refers to calling models through code to execute specific tasks, while \textit{Deployment implementation} (60$\times$) involves challenges encountered when deploying models locally or on servers. 
Some discussions explicitly mention deployment environments, including \textit{Docker} (8$\times$, containerization platform), \textit{AWS SageMaker} (6$\times$, cloud-based machine learning service), and \textit{Google Colab} (5$\times$, cloud-based notebook environment). 

(3) \textit{\textbf{Training \& Fine-tuning}} covers discussions on adapting and optimizing models for specific tasks. 
\textit{Fine-tuning} (49$\times$) is the most frequently discussed topic, indicating strong demand for task-specific performance improvement. 
\textit{LoRA/QLoRA} (13$\times$), widely used parameter-efficient fine-tuning techniques, are frequently mentioned (11$\times$ from MLLM users). 
\textit{Training data} (7$\times$) involves inquiries about datasets used for model training, such as requests for access to pretraining data. 
\textit{Training parameters} (4$\times$) refer to questions about configuration settings during training or fine-tuning. 
\textit{Model weights} (5$\times$) include issues related to weight loading as well as related inquiries. 
\textit{Training loss monitoring} (2$\times$) captures abnormal behaviors observed during training loss tracking.

(4) \textit{\textbf{Access \& Licensing}} involves discussions on model access control, licensing terms, and citation requirements.  
\textit{Access permission} (85$\times$) is the most frequently discussed topic, mainly concerning requests for model access (69$\times$ from GLLM users and 16$\times$ from MLLM users). On Hugging Face, some LLMs are gated and require users to apply for access before they can download and deploy them. 
\textit{DOI requests} (47$\times$) reflect users’ needs for assigning Digital Object Identifiers (DOIs) to models to enable reliable and standardized citation in both academic research and software projects (42$\times$ from GLLM users and 5$\times$ from MLLM users).  
\textit{Licensing} (5$\times$) covers discussions on legal terms governing model usage, including whether models can be used commercially, modified, or redistributed.

\subsubsection{\textbf{Reliability}}

This category captures user discussions on model reliability, including issues related to bias, moderation, and system errors, comprising two subcategories: \textit{Bias and Moderation} and \textit{Bugs and Errors}, with four leaf nodes in total. 

(1) \textit{\textbf{Bias and Moderation}} involves discussions on model biases and content moderation mechanisms. 
\textit{Bias and fairness} (2$\times$) refers to concerns about biased or potentially unfair or discriminatory outputs.  
\textit{Content moderation} (2$\times$) captures user concerns about moderation policies, particularly cases where restrictions are perceived as overly strict (e.g., blocking mildly sensitive or borderline inappropriate content). 

(2) \textit{\textbf{Bugs and Errors}} covers system-level defects and abnormal model behaviors observed during usage.  
\textit{Abnormal outputs} (22$\times$) refer to clearly incorrect or unexpected generation results, such as garbled text or empty responses.   
\textit{Bug reports} (14$\times$) capture user-identified defects or system errors during model usage, including issues not directly reflected in output content (e.g., crashes, incorrect behaviors, or malfunctioning features).

\subsubsection{\textbf{Supportability}} 
This category captures user discussions on integrating models into external platforms and ensuring compatibility with underlying dependencies. It consists of two subcategories, namely \textit{Third-party Integration} and \textit{Library Dependencies}, with 12 leaf nodes in total. 

(1) \textit{\textbf{Third-party Integration}} focuses on deploying and integrating models within external platforms and application frameworks. Users share practical experiences of integrating models into widely used platforms and frameworks, including \textit{ComfyUI} (8$\times$, node-based visual workflow for generative models), \textit{Ollama} (2$\times$, local runtime environment for LLMs), \textit{llama.cpp} (4$\times$, lightweight C/C++ inference engine), \textit{Draw Things} (1$\times$, on-device image generation application), \textit{LightX2V} (1$\times$, text-to-video generation framework), \textit{WebUI} (1$\times$, browser-based user interface), and \textit{HuggingChat} (1$\times$, web-based chat interface provided by Hugging Face).

(2) \textit{\textbf{Library Dependencies}} concerns model adaptation to underlying libraries and system dependencies. Users report errors and compatibility issues when combining models with different software stacks, including \textit{Transformers} (20$\times$), \textit{Torch/CUDA} (4$\times$), \textit{FlashAttention} (4$\times$, optimized attention computation library), \textit{DeepSpeed} (1$\times$, distributed training and inference framework), and \textit{MLX} (1$\times$, Apple machine learning framework).

\subsubsection{\textbf{Performance}}

This category captures user discussions on runtime requirements and execution efficiency. It includes three subcategories, namely \textit{Resource Consumption}, \textit{Inference Performance}, and \textit{Model Efficiency}, with eight leaf nodes in total.

(1) \textit{\textbf{Resource Consumption}} concerns the use of computational resources and power. \textit{GPU/Memory usage} (27$\times$) is the most frequently discussed topic, including requirements for GPU memory and RAM as well as issues such as out-of-memory errors. \textit{Hardware requirements} (18$\times$) refer to deployment constraints and performance across different hardware environments (e.g., A100 GPUs). \textit{Energy consumption} (2$\times$) involves power usage and heat generation during model execution.

(2) \textit{\textbf{Inference Performance}} focuses on generation speed and batch processing capability. \textit{Inference latency} (9$\times$) refers to the time from input to output generation, while \textit{batch efficiency} (1$\times$) captures the model’s throughput in processing multiple tasks simultaneously.

(3) \textit{\textbf{Model Efficiency}} involves discussions on model optimization techniques for improving efficiency, including quantization and architectural lightweighting.  \textit{Quantization} (16$\times$) refers to reducing numerical precision (e.g., FP16 to INT8) to improve efficiency and reduce resource usage. In addition, users express demand for \textit{GGUF} (5$\times$, efficient binary format for LLM inference) and \textit{MoE} (3$\times$, mixture-of-experts architecture) variants.

\subsubsection{\textbf{Comparison}} 
This category captures user discussions on evaluations using standardized benchmarks and cross-model comparisons. It consists of two subcategories, \textit{Benchmarking} and \textit{Model Comparison}, with four leaf nodes in total.

(1) \textit{\textbf{Benchmarking}} focuses on quantitative performance on standardized public datasets. \textit{Leaderboard scores} (5$\times$) refer to model rankings on public leaderboards, while \textit{metric reproduction} (7$\times$) captures users’ efforts to reproduce reported results by running models on benchmark datasets and sharing the outcomes.

(2) \textit{\textbf{Model Comparison}} involves comparisons across different models. \textit{Version comparison} (14$\times$) refers to comparisons between different versions within the same model family, while \textit{competitor comparison} (3$\times$) captures comparisons with other models serving similar purposes.

\subsubsection{\textbf{General Experience}}

This category captures users’ overall subjective feedback and sharing when interacting with models. It contains no subcategories and consists of two leaf nodes. 
\textit{General feedback} (25$\times$) refers to users’ subjective evaluations, usage experiences, and overall impressions after interacting with the model.   
\textit{Project promotion} (26$\times$) refers to cases where users share or promote derivative work built upon the model, such as adapted models, plugins, or tools developed to enhance usability.

\subsection{RQ2: What key dimensions of LLMs do users care about most?}

\begin{table}[!t]
    \centering
    \fontsize{8pt}{8pt}\selectfont
    \caption{The top-15 most frequently discussed topics.}
    \label{tab:top-15}
    \begin{tabular}{@{}
        >{\raggedright\arraybackslash}p{0.15cm}
        |
        >{\raggedright\arraybackslash\hspace{-2pt}}p{3.2cm}
        >{\raggedleft\arraybackslash}p{0.2cm} 
        |
        >{\raggedright\arraybackslash\hspace{-2pt}}p{3.2cm}
        >{\raggedleft\arraybackslash}p{0.2cm} 
        @{}
    }
    \toprule
        &
        \multicolumn{2}{c|}{\textbf{GLLM User Discussions}} & \multicolumn{2}{c}{\textbf{MLLM User Discussions}} \\
    \midrule
         & \multicolumn{1}{c}{\textbf{Label}} & No. & \multicolumn{1}{c}{\textbf{Label}} & No. \\
    \midrule
        1  & Access permission & 69 & Generation quality & 46 \\
        2  & DOI requests & 42 & Model invocation & 40 \\
        3  & Deployment implementation & 25 & Deployment implementation & 35 \\
        4  & Model invocation & 21 & Fine-tuning & 28 \\
        5  & Fine-tuning & 21 & Documentation & 26 \\
        6  & Documentation & 17 & Example code & 17 \\
        7  & Project promotion & 15 & Access permission & 16 \\
        8  & General feedback & 15 & GPU/Memory usage & 14 \\
        9  & Prompt and template usage & 14 & Transformers & 13 \\
        10 & Abnormal outputs & 13 & Quantization & 11 \\
        11 & GPU/Memory usage & 13 & Project promotion & 11 \\
        12 & Example code & 12 & Version comparison & 11 \\
        13 & Hardware requirements & 11 & General feedback & 10 \\
        14 & Bug reports & 8 & Prompt and template usage & 9 \\
        15 & Transformers & 7 & Abnormal outputs & 9 \\
        \bottomrule
    \end{tabular}
\end{table}

Following prior studies~\cite{pearson1900x, cochran1954some, sharpe2015your}, we construct an $18 \times 2$ contingency table (rows: 17 subcategories plus \textit{General Experience}; columns: GLLM vs.\ MLLM) and perform a Pearson chi-square test. The results indicate a statistically significant difference in the distribution of discussion topics between the two groups ($\chi^2 = 96.45$, $\text{df} = 17$, $p < 0.001$), where $\chi^2$ denotes the chi-square test statistic, $\text{df}$ represents the degrees of freedom, and $p$ indicates the statistical significance of the observed difference.
To further identify the sources of these differences, we compute standardized residuals. An absolute residual value greater than 1.96 indicates that the observed frequency is significantly higher or lower than expected. The results reveal several notable differences. \textit{Functionality--Core Capabilities} is significantly overrepresented in the MLLM group (residual = +4.15), while \textit{Usability--Access \& Licensing} is significantly overrepresented in the GLLM group (residual = +8.09). In addition, \textit{Usability--Invocation \& Deployment} shows a significantly higher frequency in the MLLM group (residual = +3.13).
Then, Table~\ref{tab:top-15} presents the most frequently discussed topics (i.e., leaf nodes) for general and multimodal LLMs, along with the corresponding number of discussion threads. Based on this analysis, we derive seven key findings that highlight the dimensions users care about most. To substantiate these findings, we quote excerpts from the original discussion threads, where each thread is referenced by an identifier (e.g., (D1, GLLM) denotes the first entry in the dataset, which originates from GLLM).

\noindent \faLightbulbO \quad  \textbf{Finding 1. Access barriers, particularly in the Llama model family, constitute a primary obstacle for users interacting with LLMs.} 
\textit{Access permission} emerges as the most frequently discussed issue among GLLM users. As shown in Table~\ref{tab:top-15}, \textit{Access permission} ranks first in GLLM discussions (69 threads) but only seventh in MLLM discussions (16 threads). Similarly, \textit{DOI requests} appear 42 times in GLLM discussions but only 5 times in MLLM discussions. 
A closer examination shows that most threads related to \textit{Access permission} and \textit{DOI requests} are associated with the Llama model family. In our dataset, 7 out of the 21 selected GLLMs are Llama-based, whereas only 2 out of 17 MLLMs belong to this family. As a result, \textit{Usability--Access \& Licensing} related discussions are disproportionately concentrated in the GLLM group. 
Unlike traditional open-source models released under OSI-approved licenses (e.g., MIT, Apache 2.0, or GPL), Llama models adopt Meta’s custom Community License Agreement, which introduces additional access requirements such as application-based approval and usage constraints. These mechanisms increase the barrier to adoption and lead to frequent user concerns. Although these issues are most prevalent in the Llama model family, similar access barriers are also observed in other models. For example, multiple users report access denied errors when invoking \textit{mistralai/Mistral-7B-Instruct-v0.2} (D166, GLLM).   
User feedback further suggests that these access barriers mainly stem from opaque approval decisions and delayed or unpredictable response times. One user questions the lack of transparency in the approval process:

\noindent\faEdit \textit{They rejected me for no reason too. All they asked for was my name and affiliation, so I was either rejected because I listed myself as an ``independent researcher'' or because of my name. Neither makes any sense.} (D245, GLLM).

Other users report long and highly variable waiting times before obtaining access:

\noindent\faEdit \textit{Left a comment 4 days ago, but still no response.} (D31, GLLM).

\noindent\faEdit \textit{Normally, it would need less than 2 hours to get permission. But for llama-4 models, it's been 3 days and it's still pending.} (D466, MLLM).

\noindent \faLightbulbO \quad  \textbf{Finding 2. Generation quality is the core evaluation dimension: MLLM users are more sensitive than GLLM users, and prompt design critically shapes perceived quality.} 
Our sentiment annotation shows that most user feedback is \textit{neutral}, with only 18 threads expressing \textit{like} and 17 expressing \textit{dislike}. These sentiments are primarily concentrated in \textit{General feedback}. Within 25 \textit{General feedback} threads, 17 express \textit{like}, 4 express \textit{dislike}, and 4 are \textit{neutral}. Overall, this suggests that many users acknowledge the capabilities of current models. For example, one user comments: 

\noindent\faEdit \textit{The consistency of face and body type is incredible, not to mention the clarity.} (D513, MLLM).

However, in more complex or high-demand tasks, model outputs often fall short of user expectations due to task complexity and detailed requirements. In total, we identify 75 threads related to output-related issues, including 53 on \textit{Generation quality} and 22 on \textit{Abnormal outputs}. 
As shown in Table~\ref{tab:top-15}, \textit{Generation quality} ranks first among MLLM users (46 threads) but falls beyond the top 15 among GLLM users (7 threads). This discrepancy reflects the additional challenges MLLMs face in aligning text with visual or audio modalities, increasing the likelihood of errors:

\noindent\faEdit \textit{When generating images in the 4-step process, there will be matrix-like dot-like artifacts. They are not very obvious but can be easily noticed as there are faint matrix-like artifacts.} (D441, MLLM). 

\noindent\faEdit \textit{There are too many cases where the face changes drastically or the skin is rendered with an overly plastic appearance. } (D455, MLLM)

We also observe that generation-quality discussions are highly concentrated on a small number of models rather than being evenly distributed across the ecosystem. For example, the community-maintained MLLM \textit{Phr00t/Qwen-Image-Edit-Rapid-AIO} alone accounts for 31 generation-quality-related threads. This concentration suggests that users are particularly sensitive to output quality, with quality deficiencies quickly attracting substantial community attention even for relatively niche models. 

For GLLMs where foundational generation and reasoning are relatively mature, users focus more on output completeness, logical coherence, and robustness. For instance, one user reports inconsistent responses across instructions:

\noindent\faEdit \textit{in some cases it works and of some query it return half ans. Greetings are not handled properly it return un,matched ans} (D286, GLLM).

We further observe that \textit{prompt and template usage} play a critical role in generation quality, with 14 mentions in GLLM discussions and 9 in MLLM discussions. Well-designed prompts and templates help models better understand task requirements and produce outputs aligned with expected formats, whereas poorly designed ones can significantly degrade output quality. For example, when using a template such as \texttt{"""\{context\}\textbackslash n\textbackslash n\{question\}"""}, one user reports that the model only returns overly concise answers like \textit{``28 Kilometer"} (D120, GLLM), failing to meet task expectations.

\noindent \faLightbulbO \quad  \textbf{Finding 3. Model invocation and deployment are the most common practical challenges for developers, with MLLM users facing a steeper learning curve than GLLM users.}
Model invocation and deployment represent critical stages in model application. Users of general and multimodal LLMs frequently encounter resource adaptation issues and unstable runtimes. Even on standardized platforms, GPU memory overflow and framework conflicts remain common obstacles.
As shown in Table~\ref{tab:top-15}, MLLM users report these issues more frequently, with 75 threads discussing \textit{invocation \& deployment} compared to 46 threads for GLLMs. This likely reflects differences in ecosystem maturity: GLLMs rely on standardized toolchains, whereas emerging MLLMs involve heterogeneous architectures (e.g., integrating visual encoders with LLMs), leading to more complex and less standardized preprocessing and memory management. 
During model invocation, although Hugging Face's \texttt{pipeline()}~\cite{hf_pipeline_tutorial} interface substantially lowers the barrier to entry, we observe that relatively few discussion threads explicitly mention using this high-level interface. Specifically, among the 61 threads labeled as \textit{Model invocation}, only 5 explicitly describe invoking models through \texttt{pipeline()}. Instead, most discussions focus on manually loading and invoking models, suggesting that users often seek greater flexibility and control over the inference process. However, even when using \texttt{pipeline()}, users may encounter issues caused by the lag between model releases and framework support. For example, one user reports being unable to run the \textit{mistralai/Mistral-7B-v0.1} model because support for the model architecture had not yet been integrated into the Transformers library, resulting in execution failures despite following the recommended usage approach (D56, GLLM).  In addition, inference errors and service exceptions are common, including HTTP 503 timeouts and input format mismatches. For example, one user reported:

\noindent\faEdit \textit{Hi just wanted to use this with transformers but now as Response I always get 503 and here on the test it is said that the service is unavailable.} (D189, GLLM).

In the deployment stage, challenges shift to environment packaging and complex dependency integration. Offline loading or missing local dependencies often causes task failures, as another user notes:

\noindent\faEdit \textit{I have downloaded all the checkpoints files in my local. And when I upload the model using directory path, it does not load it.} (D61, GLLM).

MLLM users also encounter deployment issues when transitioning to platforms like Colab:

\noindent\faEdit \textit{When i use this model on the Colab , the inference is happening fine , but once I load it in local , it pops a error message called ``pip install flash\_attn".} (D371, MLLM).

\noindent \faLightbulbO \quad  \textbf{Finding 4. Insufficient and error-prone documentation and tutorials significantly hinder model adoption, increasing the learning cost for developers.}
Clear documentation and reliable example code are vital for effective model adoption, particularly during invocation and deployment. Developers depend on official guidance for data preprocessing and system configuration, especially for multimodal inputs. However, missing details—such as unspecified image preprocessing requirements—frequently lead to failures, as current documentation often falls short. Prior research ~\cite{taraghi2024deep,chang2022understanding} indicates that only about 52\% of models on Hugging Face provide complete model cards, which practitioners attribute to a combination of insufficient incentives, fragmented tooling support, and challenges in determining appropriate documentation content. User feedback further validates this observation. Beyond basic API descriptions, developers frequently seek task-oriented documentation that offers practical guidance for real-world applications. However, such contextualized resources remain scarce within the current ecosystem. For example, one user explicitly requests clearer guidance:

\noindent\faEdit \textit{How do I get streaming token generation from mistral\_common? Example needed.} (D273, GLLM).

In addition, errors and inconsistencies in documentation and example code further increase the burden on developers. For instance:

\noindent\faEdit \textit{Example code in README.md doesn't work.} (D215, GLLM).

\noindent\faEdit \textit{I ran the Using Pipeline code exactly as it is in the README, but I'm getting a dimension error. Additionally, there is a typo (import request -$>$ import requests).} (D522, MLLM).

\noindent \faLightbulbO \quad  \textbf{Finding 5: Fine-tuning is widely demanded but remains technically challenging; its adoption is particularly prominent for smaller models, where parameter-efficient methods still face compatibility and performance issues.} Although base models exhibit strong generalization, their performance in specific domains often motivates developers to pursue fine-tuning. Table~\ref{tab:top-15} shows high interest in this area, with 21 threads for GLLMs and 28 threads for MLLMs discussing \textit{Fine-tuning}. However, fine-tuning remains technically challenging for many developers. Approximately one-fifth of the fine-tuning-related threads involve users seeking help with the fine-tuning process itself, suggesting that accessible onboarding resources and practical guidance are often lacking. These discussions cover a broad spectrum of customization needs, ranging from language adaptation to task-specific enhancement. 
For example, developers seek to adapt models to new languages:

\noindent\faEdit \textit{Since Korean is not supported in the ``3.1" Instruct model(Meta-Llama-3.1-8B-Instruct), is it recommended to use the ``3.1 basemodel"\\(Meta-Llama-3.1-8B) for fine-tuning with a Korean dataset?} (D103, GLLM).

Another user targets performance enhancements for specialized tasks:

\noindent\faEdit \textit{The model has shown commendable results on OCR for printed text, but for handwritten text, it faces some challenges. How can we fine-tune the model for OCR, especially for handwritten text?} (D637, MLLM).

In practice, the technical challenges of fine-tuning have not disappeared with the emergence of parameter-efficient methods. Although LoRA substantially lowers hardware and computational requirements, it introduces new challenges related to performance degradation and framework compatibility. For example, one user reports:

\noindent\faEdit \textit{Using the llava-instruct-chinese dataset, the image encoder weights are frozen, and the language part of Florence-2 is fine-tuned using the LoRA method. While performing the captioning task, the model is capable of outputting in Chinese, but the accuracy of the answers is zero.} (D442, MLLM). 

Another user encounters compatibility issues when fine-tuning \textit{Idefics-8B} with LoRA, resulting in failures during the reconstruction of \texttt{gather\_map} (D609, MLLM). 

Interestingly, fine-tuning discussions are more concentrated around smaller models. For example, Microsoft’s Phi-2 (2.7B) and Florence-2-large (0.77B) see 25.0\% and 27.3\% of their discussion tags related to fine-tuning, respectively. A similar size-driven pattern emerges within Google’s Gemma family: the smaller Gemma-3-4B-it and Gemma-7B each sit at 12.5\%, whereas the larger Gemma-3-27B-it accounts for only 4.0\%.  This suggests that fine-tuning demand is not solely driven by model capability, but also by practical considerations: smaller models impose fewer computational barriers and are therefore more accessible for customization and experimentation.

Beyond computational constraints, inconsistent fine-tuning workflows further increase development costs. Although \texttt{chat\_template} provides a standardized mechanism for prompt formatting, many community tutorials still rely on manually defined templates, causing confusion between different data formats. For example, one user reports difficulties fine-tuning Llama 3.1 due to inconsistent formatting practices across online tutorials (D125, GLLM).

\noindent \faLightbulbO \quad  \textbf{Finding 6. GPU and system memory limitations constrain local deployment, and quantization techniques are used to alleviate these constraints.}
As model sizes grow, developers often face conflicts with limited hardware resources. GPU memory overflow is a common obstacle during deployment and inference, affecting both consumer-grade and professional hardware. As shown in Table~\ref{tab:top-15}, \textit{GPU/Memory usage} issues appear 13 times for GLLMs and 14 times for MLLMs. 
For example, one user reports difficulty running a high-parameter model on dual 3090 GPUs:

\noindent\faEdit \textit{I am 2x3090 User. I can run Lama 30b model. in fact i can open Lama 65b but cant run cuz of memory (system memory not cuz of vram).} (D9, GLLM).

In the multimodal domain, additional overhead from image encoding increases memory demands, posing greater challenges for entry-level GPUs:

\noindent\faEdit \textit{I use the following code and I receive the error ``OutOfMemoryError: CUDA out of memory. Tried to allocate 112.00 MiB. GPU". I have nvidia 3050. what can be the problem?} (D646, MLLM).

To mitigate these hardware constraints, users increasingly adopt quantization techniques to reduce resource requirements. As shown in Table~\ref{tab:top-15}, \textit{Quantization} appears 11 times among MLLM discussions, making it one of the frequently discussed issues. In addition, we identify 5 related discussions among GLLM users, resulting in 16 quantization-related threads in total. These discussions mainly focus on weight quantization, particularly 8-bit and 4-bit solutions such as bitsandbytes (3/16), GPTQ (1/16), and GGUF (6/16). In contrast, ultra-low-precision methods (e.g., 2-bit or lower) receive little attention in user discussions. By compressing model weights to lower-bit representations, quantization techniques help reduce memory consumption and computational costs while maintaining acceptable performance. For example, one user reports: 

\noindent\faEdit \textit{I faced a CUDA memory issue when running \texttt{run\_fuyu\_no\_trainer.py} on a single A100 card. I resolved this by using 4-bit models, which happen to work on either a 3090 or a A100 card.} (D459, MLLM).

In addition, community members actively share their own quantization solutions. For instance, a user provides a quantized version of \textit{meta-llama/Llama-3.1-8B-Instruct}, highlighting its reduced model size while preserving performance (D1, GLLM).

\noindent \faLightbulbO \quad  \textbf{Finding 7. Although benchmark evaluations are widely used for model comparison, users frequently encounter difficulties reproducing reported results and often question the reliability of benchmark scores.} 
As the number of LLMs continues to grow rapidly, benchmark evaluations have become an important reference for model selection and cross-model comparison. Scores reported on leaderboards, including general-purpose benchmarks such as MMLU-Pro~\cite{wang2024mmlu} and specialized evaluations such as SWE-bench~\cite{jimenez2024swe} and MATH-500~\cite{lightman2024let}, can substantially influence community perceptions of model capabilities and adoption decisions. However, among the 12 benchmark-related discussions under the \textit{Comparison} category, approximately half involve difficulties reproducing reported metrics. Users who evaluate models using officially recommended frameworks and configurations often obtain results that differ noticeably from those reported in model cards or leaderboards. For example, two users noted:  

\noindent\faEdit \textit{I can't get an accuracy of 68.4 or close to it using opencompass, just 66.64. (D93, GLLM)}

\noindent\faEdit \textit{Can you share the script to test the results of Fuyu-8B on the OKVQA dataset? I used the prompt of ``Answer the following OKVQA question based on the image: {}", and I got a result of 20.2. (D396, MLLM)}


Beyond reproducibility issues, users also question whether benchmark scores fully reflect practical model capabilities. For example, one user comments: 

\noindent\faEdit \textit{Qwen3 235b is a grossly overfit mess. Normally when you train on a larger corpus, in this case over 30 trillion tokens, broad knowledge and SimpleQA scores go up per parameter count. But Alibaba so grossly overfit coding, math, and STEM that its much older and smaller Qwen2 72b has vastly more broad knowledge and a notably higher SimpleQA score ($\sim$18). (D124, GLLM)}

Collectively, these discussions suggest that while benchmark scores remain an important reference for model selection, users increasingly expect evaluations to better reflect practical capabilities, robustness, and alignment with real-world usage scenarios. 

\noindent \faLightbulbO \quad  \textbf{Other Findings.} 
Beyond the primary findings, we identify several additional insights from Table~\ref{tab:top-15}. First, \textit{Version comparison} emerges as a notable concern among MLLM users, ranking 12th among the top discussion topics, while it does not appear in the top 15 for GLLM users. This suggests that MLLM users are more attentive to model updates and performance differences across versions. A plausible explanation is that multimodal capabilities (e.g., image generation and visual understanding) produce more perceptible outputs, making performance variations across versions easier to observe and evaluate. 
Second, MLLMs exhibit more integration issues with the Transformers library. Discussions related to \textit{Transformers} rank 9th among MLLM topics but do not appear in the top 15 for GLLMs. This discrepancy likely stems from MLLM architectural complexity involving heterogeneous components like encoders and fusion modules. Unlike GLLMs, MLLMs face immature support within ecosystems like Transformers, exacerbating compatibility challenges. Third, substantial differences exist in the sentiment composition of discussions across model providers. DeepSeek exhibits the highest like-to-dislike ratio among all providers (8:1), and 29\% of its discussions focus on overall user experience, compared to only 2\%–9\% for other providers. This pattern may be attributable to DeepSeek’s combination of strong performance, low cost, and openness, which encourages users to evaluate and discuss the model from a holistic usage perspective rather than focusing solely on specific technical issues. Fourth, although multilingual support does not rank among the most frequently discussed topics, it reflects a critical requirement for global deployment. Users consistently express expectations for stronger support of non-English languages, particularly Asian languages. Requests range from pretrained versions tailored to French (D22, GLLM) to explicit appeals such as ``Please support Asian languages such as Japanese, Chinese, and Korean" (D567, MLLM). These discussions suggest that multilingual capability is not merely an optional feature but a fundamental prerequisite for the global adoption of LLMs.

\section{Implications}
\label{sec:implications}

Our findings reveal a broader shift in the LLM ecosystem: while previous efforts have primarily focused on improving model capabilities, real-world adoption is increasingly constrained by engineering usability, including accessibility, deployment complexity, documentation quality, resource efficiency, and evaluation reliability. Based on these observations, we derive implications for different stakeholders.

\noindent \textbf{Implications for Model Developers and Providers.} 
(1) Model development should move beyond capability-centric optimization toward usability-oriented engineering. Access restrictions, deployment difficulties, documentation limitations, and hardware constraints frequently prevent effective model adoption. Therefore, providers should treat accessibility and engineering support as essential dimensions of model quality.
(2) Providers should strengthen deployment infrastructure and developer support, particularly for emerging MLLMs. Their heterogeneous architectures introduce additional compatibility challenges. Standardized environments, dependency specifications, validated examples, and troubleshooting resources can reduce deployment overhead. Prior studies also identify insufficient standardization as a major obstacle to LLM adoption~\cite{hou2025unveiling}. 
(3) Documentation should evolve from static model descriptions toward task-oriented developer guidance. Incomplete model cards and unreliable examples increase learning costs. Previous studies report limitations in Hugging Face model cards~\cite{suryani2025model} and highlight the importance of high-quality documentation for software usability~\cite{barton2022make}. Providers should offer practical guidance for inference, fine-tuning, data formatting, and framework integration. 
(4) Providers should better support efficient adaptation and resource-constrained deployment. The demand for fine-tuning and quantization demonstrates the importance of model adaptability and efficiency. Providers should release optimized model variants and officially supported compression solutions, such as GPTQ~\cite{frantar2022gptq} and AWQ~\cite{lin2024awq}, to reduce deployment barriers. 
(5) Model evolution should incorporate user-oriented evaluation and global applicability. Users care not only about benchmark performance but also about generation quality, multilingual capability, and practical reliability. Providers should complement benchmark evaluations with real-world scenarios and human feedback while improving multilingual support.

\noindent \textbf{Implications for Model Users and Practitioners.} 
(1) Effective LLM adoption requires considering engineering constraints and application requirements in addition to model capability. Practitioners should leverage mature toolchains, official resources, and community knowledge to reduce deployment and debugging costs. 
(2) Practitioners should consider prompt engineering and parameter-efficient fine-tuning before adopting larger models. Model adaptability can be as important as model scale, especially when computational resources are limited. 
(3) Model selection should consider deployment environments, resource availability, and reproducibility rather than relying solely on benchmark rankings. Application-specific evaluation and human judgment provide complementary evidence for model selection.

\noindent \textbf{Implications for Researchers.} 
(1) LLM evaluation methodologies should incorporate real-world usage scenarios, engineering effort, and user perceptions beyond traditional benchmarks. Future research should develop evaluation frameworks that better characterize practical model quality and adoption experience. 
(2) Software engineering challenges surrounding LLM deployment require further investigation. Dependency management, framework compatibility, environment configuration, and resource optimization remain important research opportunities for improving LLM reliability. 
(3) Efficient model adaptation and compression remain important directions. Since current discussions mainly focus on 4-bit and 8-bit quantization, future research could explore high-fidelity optimization techniques that reduce resource requirements while preserving practical model quality.

\section{Threats to Validity}
\label{sec: threats to validity}

\textit{Internal Validity.} The manual annotation and taxonomy construction introduce potential bias. Although two experienced annotators conduct the labeling with multiple rounds of discussion and arbitration by an additional researcher, achieving high inter-rater agreement (Cohen’s $\kappa > 0.87$), their prior knowledge may still introduce confirmation bias, particularly when interpreting discussions on emerging MLLMs. In addition, the three-level taxonomy (category–subcategory–leaf node) is iteratively developed, and the initial pilot annotation of 200 threads introduces a potential anchoring effect, which may constrain the emergence of new categories despite iterative refinement. 
In addition, we do not systematically analyze the evolution of user concerns across model versions, which may limit the scope of our findings. Among the 38 models included in our dataset, the DeepSeek, Qwen, and Llama families contain a relatively large number of model variants. However, the discussion volume for individual DeepSeek and Qwen versions is insufficient to support reliable version-level analysis. For the Llama family, which provides richer version-specific discussions, we observe that discussions across versions are consistently concentrated on \textit{Access \& Licensing} and \textit{Invocation \& Deployment}, with limited variation in topic distribution. The two annotators further review all threads under these dominant themes and do not identify additional recurring concerns or version-specific patterns beyond those reported in our study. Therefore, we believe that the absence of version evolution analysis is unlikely to substantially affect our main findings, although future studies with more extensive longitudinal data may reveal evolutionary trends that are not captured in our dataset.

\textit{External Validity.} Our dataset is limited to Hugging Face and includes only 38 highly active models (21 GLLMs and 17 MLLMs, each with at least 50 discussion threads), which limits generalizability to other platforms (e.g., ModelScope, PyTorch Hub) or proprietary models (e.g., GPT-4, Claude). Models with limited popularity often have too few discussion threads to support meaningful qualitative analysis. As a result, our study is restricted to models with sufficiently active communities. Consequently, the findings may not fully capture issues specific to niche or emerging models. Although 662 threads are randomly sampled with a 95\% confidence level and 5\% margin of error, the sample may not fully represent the broader LLM ecosystem. Nevertheless, the scale of the dataset is comparable to that of prior empirical studies on LLMs~\cite{lyu2025my, wang2024understanding, yu2026does}. 
Furthermore, the removal of extremely short or image-only posts excludes some modality-specific feedback.

\textit{Construct Validity.} The operationalization of user concerns relies on the proposed taxonomy, which, despite iterative refinement into 7 top-level categories and 17 subcategories, still involves ambiguous boundaries. For example, distinctions between \textit{Generation Quality} and \textit{Reasoning Ability} may depend on subjective interpretation. In addition, sentiment labeling (like/dislike/neutral) is conservative and is only applied when explicit opinions are expressed, which may underestimate nuanced or implicit user dissatisfaction.

\textit{Conclusion Validity.} Our findings are supported by qualitative evidence from selected discussion excerpts. However, this process introduces potential selection bias, as examples that align with observed patterns are more likely to be reported, potentially overlooking contradictory cases and leading to partial interpretations of user needs.

\section{Related Work}
\label{sec:related work}

\begin{table*}[t]
    \centering
    \fontsize{8pt}{9.5pt}\selectfont 
    \setlength{\tabcolsep}{2pt}
    \caption{The overview of differences in data sources, models, and finding focuses between our study and related studies.}
    \label{tab:related_studies}
    \begin{tabular}{@{}l | l | l | m{12.6cm}@{}}
    \toprule
    \textbf{Study} & \textbf{Data Source} & \textbf{Models} & \textbf{Focus Areas of Findings} \\
    \midrule
    Katta~\cite{katta2025analyzing} & Reddit & \makecell[l]{ChatGPT, \\ DeepSeek} &  development and training; practical application and user experience; legal and ethical issues    \\
    \midrule
    Wang~\cite{wang2024understanding} & Survey & GLLMs & usage patterns and intents; scenario diversity; user experience; user expectation gaps \\
    \midrule
    Lyu~\cite{lyu2025my} & \makecell[l]{VS Code \\ Marketplace}  & Code assistant & productivity perception; generation quality; context awareness limitations; usability, resource usage, pricing \\
    \midrule
    Yu~\cite{yu2026does} & GitHub Issues & 
    \makecell[l]{DeepSeek,\\Llama, Qwen} & runtime instability and infrastructure failures; dependency conflicts; user configuration faults; generation quality; quantization\\
    \midrule 
    Chen~\cite{chen2025empirical} & OpenAI Forum & GPT & LLM development; API usage issues; integration with applications; prompt design; non-functional constraints \\  
    \midrule 
    Our Study & Hugging Face & \makecell[l]{GLLMs, \\ MLLMs}  & \textbf{access and licensing barriers}; generation quality; \textbf{deployment and invocation challenges}; \textbf{documentation limitations}; fine-tuning needs; GPU and system memory limitations; quantization;  \textbf{version comparison and benchmarking}; library dependencies; \textbf{multilingual support} \\
    \bottomrule
    \end{tabular}
\end{table*}

\noindent \textbf{\textit{User Perceptions and Developer Challenges of LLMs.}} Katta~\cite{katta2025analyzing} analyzed Reddit discussions on ChatGPT and DeepSeek using sentiment analysis and LDA topic modeling, identifying three topics: LLM development and training, LLM applications and user interaction, and legal and ethical concerns. Wang et al.~\cite{wang2024understanding} conducted a survey to understand users’ experiences with LLMs, focusing on usage scenarios, satisfaction, preferences, and expectations for future tools. 
Some studies further investigate user perceptions of LLMs in specific application contexts. Bernabei et al.~\cite{bernabei2023students} conducted a case study to investigate students’ use of ChatGPT for academic writing and their perceptions of LLMs in education. Choudhury et al.~\cite{choudhury2025user} analyzed 556 survey responses on user perceptions of DeepSeek in healthcare scenarios and found that users are generally receptive to DeepSeek. In the context of LLM-based coding assistants, several studies~\cite{github2024survey,liang2024large,sergeyuk2025using,wang2023practitioners,mcnutt2023design,ziegler2022productivity} have used surveys and interviews to examine users’ experiences, interaction patterns, and expectations. Lyu et al.~\cite{lyu2025my} analyzed 361 user reviews from the VS Code Marketplace to explore developers’ perceptions of LLM-based coding assistants, identifying key needs, benefits, and pain points. 
While these studies provide valuable insights into user perceptions of LLMs or LLM-based applications, they primarily focus on \textbf{end users} of LLM-based applications. Moreover, survey-based approaches (e.g.,~\cite{wang2024understanding,choudhury2025user}) rely on predefined questions, which may limit the diversity of feedback captured from real-world usage.

In contrast to end-user perspectives, several studies examine the challenges faced by \textbf{developers} when integrating and deploying LLMs in real-world systems. 
Yu et al.~\cite{yu2026does} analyzed 705 GitHub issues related to DeepSeek, Llama, and Qwen to investigate failure symptoms, root causes, and their relationships during the fine-tuning and deployment of open-source LLMs. Chen et al.~\cite{chen2025empirical} analyzed 2,364 posts from the OpenAI developer forum to identify challenges encountered by developers when building applications with GPT models, such as API usage and system integration. However, prior studies typically focus on a limited set of LLMs and examine either end-user experiences or developer challenges during deployment, overlooking broader community feedback across diverse models. They also rarely consider emerging MLLMs, despite their growing importance in the LLM ecosystem. In contrast, we study the LLM ecosystem as a whole rather than individual models.

Table \ref{tab:related_studies} compares our study with five most closely related works. The final column presents the core focus areas of the findings reported in these studies. The bolded items highlight aspects that have received limited attention in prior studies.  
Specifically, our study uncovers several previously underexplored aspects, including access and licensing barriers as primary obstacles to model adoption, practical challenges in model invocation and deployment in real-world settings, and the impact of insufficient or error-prone documentation on developers’ learning costs. In addition, we identify emerging concerns related to version comparison and benchmarking across different models, as well as the growing demand for multilingual support.

\noindent \textbf{\textit{Empirical Studies on Hugging Face.}} 
Hugging Face has become a central platform for hosting and sharing machine learning models, and several studies have examined its ecosystem from different perspectives. Ait et al.~\cite{ait2024hfcommunity,ait2025suitability} developed the HFCommunity tool to mine Hugging Face repositories and evaluated the platform’s reliability as a research data source, while Suryani et al.~\cite{suryani2025model} constructed a dataset of Hugging Face model card metadata. 
Research has also explored the evolution and characteristics of models on the platform. Castaño et al.~\cite{castano2024analyzing} analyzed the growth and maintenance activities of Hugging Face models, reporting exponential model growth and showing that maintenance is primarily perfective. Laufer et al.~\cite{laufer2025anatomy} studied model evolution from an evolutionary biology perspective and found strong family resemblance among models, where sibling models are often more similar than parent–child models. Ajibode et al.~\cite{ajibode2025towards} further highlighted inconsistencies and the lack of standardization in model naming. Castaño et al.~\cite{castano2026machine} found that model maintenance on Hugging Face is dominated by output data and metadata management rather than code changes, with model weight modifications declining relative to documentation and configuration updates in later lifecycle stages. Chen et al.~\cite{chen2026first} further constructed an AI supply chain covering 1.82 million models and found that 96\% of models contain at least one missing or incorrect critical metadata field, while operations such as fine-tuning and quantization can significantly affect risk propagation.

In the context of model reuse and documentation, Jiang et al.~\cite{jiang2023empirical} performed an empirical study of pretrained model reuse and proposed a reuse decision workflow. Taraghi et al.~\cite{taraghi2024deep} analyzed Hugging Face forum discussions and identified 17 categories of challenges associated with model reuse. Yang et al.~\cite{yang2025ecosystem} constructed the LLM4Code ecosystem dataset and reported the growing adoption of lightweight reuse techniques, such as quantization and format conversion, as well as the absence of license information in more than 60\% of models. Regarding documentation transparency, Pepe et al.~\cite{pepe2024hugging} analyzed 159,000 Hugging Face models and found that only 14\% disclosed training datasets, 18\% reported bias information, and 32\% specified licenses. Bhat et al.~\cite{bhat2023aspirations} further showed a substantial gap between model card guidelines and their practical adoption, reporting that Hugging Face exhibits the poorest documentation quality among major platforms and that fewer than 10\% of models provide ethical information. Other work examines security and sustainability aspects of the ecosystem. Kathikar et al.~\cite{kathikar2023assessing} identified critical vulnerabilities in open-source AI repositories and showed that they can propagate to downstream repositories. Castaño et al.~\cite{castano2024lessons,castano2023exploring} investigated the environmental impact of models and reported low and non-standardized carbon emission disclosures on Hugging Face. In addition, González et al. \cite{gonzalez2025cataloguing} proposed an automated cataloging framework that aligns model descriptions with software engineering tasks. Despite these efforts, there is still a lack of studies that systematically analyze user comments on Hugging Face to understand community perceptions of both GLLMs and MLLMs.

\section{Conclusions and Future Work}
\label{sec:conclusion}

This paper presents an empirical analysis of user discussions on Hugging Face to understand how GLLMs and MLLMs are perceived, adopted, and challenged in real-world usage. Our findings reveal that the primary barriers to LLM adoption are shifting beyond model capability toward ecosystem usability. Access and licensing barriers, particularly in the Llama model family, remain a major obstacle for users, indicating that model openness does not necessarily guarantee practical accessibility. Generation quality emerges as a central dimension of user evaluation, with MLLM users being particularly sensitive to output quality and prompt design. Meanwhile, model invocation and deployment represent the most common engineering challenges, especially for MLLMs with more complex architectures. Fine-tuning is widely adopted for customization, particularly for smaller models, yet remains technically challenging even with parameter-efficient methods such as LoRA. In addition, insufficient documentation increases developer learning costs, while hardware limitations motivate the adoption of quantization techniques. Although benchmark scores remain important references for model selection, difficulties in reproducing reported results raise concerns about the reliability and interpretability of leaderboard-based comparisons. 
For future work, we will extend our analysis to additional platforms to improve coverage and generalizability. 

\section{Data Availability Statement} 
Replication package is available at 
~\cite{replicationpackage}.

\bibliographystyle{IEEEtran}
\bibliography{sample-base}

\end{document}